\documentclass[prl,aps,twocolumn,superscriptaddress,floatfix]{revtex4-1}

\usepackage{graphicx}
\usepackage{amsmath}
\usepackage{amssymb}
\usepackage{wasysym}

\newcommand{\bra}[1]{\langle #1|}
\newcommand{\ket}[1]{|#1\rangle}
\newcommand{\expectation}[1]{\langle #1\rangle}

\newcommand{\rdm}{\rho}
\newcommand{\rdmperturb}{\tilde{\rdm}}
\newcommand{\lambdathermal}{\lambda_{\rm th}}
\newcommand{\lambdaexp}{\lambda_{\rm exp}}
\newcommand{\lambdaGauss}{\lambda_{\rm Gauss}}
\newcommand{\gammaFGR}{\gamma_{\rm FGR}}
\newcommand{\Gammashort}{\Gamma_1}
\newcommand{\Teff}{T_{\text{eff}}}

\begin{document}

\title{Dynamics of thermalisation in small Hubbard-model systems}

\author{S.~Genway}\affiliation{Blackett Laboratory, Imperial College London, London SW7 2AZ, United Kingdom}

\author{A.~F.~Ho}\affiliation{Department of Physics, Royal Holloway University of London, Egham, Surrey TW20 0EX, United Kingdom}

\author{D.~K.~K.~Lee}\affiliation{Blackett Laboratory, Imperial College London, London SW7 2AZ, United Kingdom}


\date{\today}

\begin{abstract} 
We study numerically the thermalisation and temporal evolution of 
a two-site subsystem of a fermionic Hubbard
model prepared far from equilibrium at a definite energy. Even for
very small systems near quantum degeneracy, the subsystem can reach a
steady state resembling equilibrium.  This occurs for a
non-perturbative coupling between the subsystem and the rest of the
lattice where relaxation to equilibrium is Gaussian in time, in sharp
contrast to perturbative results.  We find similar results for random
couplings, suggesting such behaviour is generic for small systems.
\end{abstract}

\maketitle

Understanding the origin of statistical mechanics from a purely
quantum-mechanical description is an interesting area of active
research~\cite{canontyp, popescu1, numerics_Jensen1985,*numerics_Saito1996, numerics_Henrich2005, 
numerics_yuan2009,rigol2008,*rigol2007, tasaki, reimann, srednicki1}.  Of particular
interest is the situation of an isolated quantum system partitioned
into a subsystem and a bath.  We ask the question: how do observables
on the subsystem thermalise when the total system is in a pure state?
Seminal works~\cite{canontyp, popescu1} have demonstrated the concept
of `canonical typicality' that most random pure states of well-defined
energy for the total system lead to thermalised reduced density
matrices (RDMs) for the small subsystem.  Numerical works have
demonstrated thermalisation in spin or boson systems for various
observables of the subsystem~\cite{numerics_Jensen1985,*numerics_Saito1996,numerics_Henrich2005, numerics_yuan2009}
(and of the entire closed system~\cite{rigol2008,*rigol2007,brody_ergodic2007,*fine2009}). Recent theoretical
work~\cite{popescu2} has investigated whether thermalisation of small
subsystems, initially far from equilibrium, is generic.

In this letter, we investigate the temporal relaxation towards a
steady state, focusing on the regime where the steady state appears
thermalised. We consider a small Hubbard ring of fermions away from
half filling, with two adjacent sites as a subsystem and the other sites
as a bath.  We prepare the system in a product state of
subsystem and bath pure states in a narrow energy window. Even for
such a small system, we find a steady-state RDM close to a thermal
state, down to quantum degenerate temperatures.
Moreover, we find that the RDM diagonal elements approach
a steady state as an exponential decay for weak subsystem-bath
coupling. This becomes a Gaussian decay at a non-perturbative
coupling, with a decay rate that departs significantly from the Fermi
golden rule. 
We note that this is distinct from the Gaussian behaviour in driven systems that remain out of equilibrium~\cite{segal2007}, and in decoherence dynamics~\cite{rossini2007,*cucchietti2007,*dobrovitski2008,*numerics_yuan2008} of off-diagonal RDM elements in systems that cannot thermalise.

\emph{The Model.}  Taking motivation from cold atoms in optical
lattices~\cite{sherson2010,bakr2010} where local addressing is possible, 
we study a local cluster in a generic (non-integrable)
interacting system with a quasi-continuous spectrum. We will examine how a
local subsystem ($S$) thermalises with the rest of the system as a bath ($B$) 
\emph{via} unitary evolution of the
whole system under the Hamiltonian 
${H} = {H}_S + {H}_B + \lambda {V}$ where ${H}_S$ (${H}_B$), with eigenstates
$\ket{s}_S$ ($\ket{b}_B$) of energy $\varepsilon_s$ ($\epsilon_b$),
acts solely on the subsystem (bath).  $\lambda V$ is the coupling
between the subsystem and the bath.  For $\lambda=0$, the eigenstates
are products of subsystem and bath eigenstates, denoted
$\ket{sb}$, with energies $E_{sb} = \varepsilon_s + \epsilon_b$. The
homogeneous case corresponds to $\lambda=1$. Specifically, we choose a
two-site subsystem in an $L$-site Hubbard ring of fermions:
\begin{eqnarray}
 {H}_S &=& 
 - \sum_{\sigma=\uparrow,\downarrow} 
 J_{\sigma} (c^{\dagger}_{1 \sigma} c_{2 \sigma} + \text{h.c.}) 
 +  U( n_{1 \uparrow} n_{1 \downarrow} +  n_{2 \uparrow} n_{2 \downarrow} ) \, , 
\nonumber\\
{H}_B &=& - \sum_{i=3}^{L-1}
\sum_{\sigma=\uparrow,\downarrow} J_{\sigma} (c^{\dagger}_{i \sigma} c_{i+1, \sigma} + \text{h.c.}) +  U \sum_{i=3}^L n_{i \uparrow} n_{i \downarrow} \, , 
\nonumber\\
{V} &=& -\sum_{\sigma=\uparrow,\downarrow} J_{\sigma}\left[ (c_{2 \sigma}^{\dagger} c_{3 \sigma} + c_{1 \sigma}^{\dagger} c_{L\sigma}) + \text{h.c.}\right]\,,
\label{eq:2linkH}
\end{eqnarray}
where $n_{i\sigma} = c^{\dagger}_{i \sigma} c_{i \sigma}$ is the
number operator on site $i$ with spin $\sigma$. The lattice 
is a ring with the subsystem sites at $i=1,2$ and bath
sites at $i=3$ to $L$. The hopping integrals are $J_{\sigma}= J(1+\xi
\; \text{sgn}(\sigma))$, with a non-zero $\xi=0.05$
to remove level degeneracies due to 
spin rotation symmetry. We set the on-site repulsion
$U=J$ to give us a metallic system with interacting bath
modes while avoiding the formation of strong features in the many-body density of states at $U\gg J$ arising from Hubbard interactions.
We will let $J$ be the unit of energy. 
This Hamiltonian preserves the total particle
number, $N$, and spin component, $S^z$, but not the total spin
$S^2$. In this work, we fill the $L=9$ lattice with 8
fermions of total spin $S^z=0$. The two-site subsystem has 16
eigenstates and the bath has 8281 eigenstates, while the composite
system has a total of 15876 states with average level spacing $\Delta \simeq 10^{-3}$. 
This is small enough to allow exact diagonalisation, but large enough to provide a smooth density of
states.

Consider a system prepared in a
pure state of the form
\begin{equation}
\label{eq:windowstate}
\ket{\Psi(t=0),E_0} =  \sum_{b_i = b_l}^{b_u} \frac{1}{\sqrt{B}} \ket{s_i b_i}
\end{equation}
where $\ket{s_i}_S$ is the initial subsystem state, \emph{e.g.} $\ket{\!\uparrow,\uparrow}_S$ with parallel spins on the two sites.  $\ket{\Psi(t=0)}$ contains a linear combination of $B$ bath eigenstates $\ket{b_i}_B$ within an energy shell of width $\delta_B$, chosen such that $\bra{\Psi}H\ket{\Psi} = E_0$. The width $\delta_B$ (= 0.5 in this work) is small on the scale for variations in the density of states.
The system evolves in time: $\ket{\Psi(t)}=e^{-iHt}\ket{\Psi(0)}$. The subsystem is described completely by the RDM which traces over the bath states $\ket{b}_B$:
$\rdm(t)= \text{Tr}_B \ket{\Psi(t)}\bra{\Psi(t)}$. This is evaluated using the eigenstates of $H$ from exact diagonalisation.

\emph{Equilibrium States.} Before discussing relaxation dynamics
during thermalisation, we identify first the parameter regime where
$\rdm$ does relax to thermal equilibrium. 
We say that a subsystem thermalises if its RDM
$\rdm(t) $ approaches the thermal RDM $\omega$ after some time $t$ 
(shorter than $\Delta^{-1}$.)
The thermal RDM, $\omega$, is diagonal with elements $\omega_{ss} =
{}_S\bra{s}\omega\ket{s}_S \propto N_B(E_0 - \varepsilon_s,
N-n_s,S^z-s^z_s)$, where $\ket{s}_S$ is a subsystem eigenstate
$\ket{s}_S$ with energy $\varepsilon_s$, $n_s$ particles and spin
$s_s^z$ and $N_B(E, n_b, s^z_b)$ is the number of bath states with
energies in $[E,E+\delta E]$ with $n_b$ particles and spin $s^z_b$. We
have to specify energy, number and spin because they are globally
conserved by the Hamiltonian $H$. We can define an
effective temperature $\Teff = [\partial \log
N_B/\partial E|_{E_0}]^{-1}$ provided that the system is in a state with energy uncertainty $\delta E \gg \Delta$, the level spacing.
(In the thermodynamic limit, $\omega$ takes the form of a Gibbs
canonical distribution~\cite{brody_ergodic2007,*fine2009} --- 
if particles are not exchanged with the bath,
$\omega_{ss} \propto e^{- \varepsilon_s/T}$.)
  
\begin{figure}[bht]
\begin{center}
\includegraphics[scale=1.25]{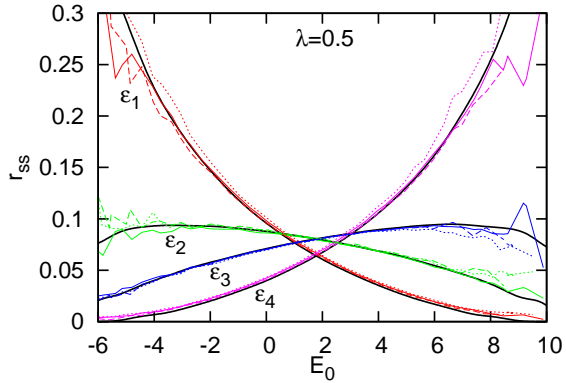}
\caption{(Colour online) Time average of diagonal RDM elements,
$r_{ss}$, for the $n_s=2$, $s^z_s=0$ sector as a function of composite
energy $E_0$, for initial subsystem states $\ket{\!\uparrow\downarrow,
\uparrow}_S$ (solid line), $\ket{\!\uparrow,\uparrow}_S$ (dashed) and
$\ket{\!\uparrow,\downarrow}_S$ (dotted).  The four elements are
labelled by their energies, ascending from $\varepsilon_1$ to
$\varepsilon_4$.  Thick lines: the corresponding elements for the
thermal state $\omega$, found by counting bath states with spin
$S^z-s^z_s$ and $N - n_s$ particles within a Gaussian energy window of
width $\delta E=0.5$, centered on energy $E_0 - \varepsilon_s$.}
\label{timeave}
\end{center}
\end{figure}
We now present our results for a system starting from the initial
states~\eqref{eq:windowstate}.  We avoid the regime of very small
subsystem-bath coupling $\lambda$ where the subsystem RDM, $\rho$, is
strongly dependent on the initial state even at long times due to
finite-size effects.  Nevertheless, we find that even such a small
system can reach a steady state for couplings $\lambda$ larger than a
surprisingly small crossover value $\lambdathermal\ll 1$. The RDM
becomes virtually diagonal --- even the sum over the fluctuating
off-diagonal elements, $[\sum_{s\neq s'} r_{ss'}^2]^{1/2}$, is
10$^{-1}$ to 10$^{-3}$ smaller than each diagonal element.
Fig.~\ref{timeave} shows the steady-state values of the diagonal
elements of the RDM, $r = \lim_{\tau \rightarrow \infty}
\int_{0}^{\tau} \rdm(t) dt/\tau$ as a function of the composite energy
$E_0$ for a coupling of $\lambda=0.5$.  For a variety of initial
states, $\rdm(t)$ is only
weakly dependent on the details of the initial state at long times 
for $-3 \apprle E_0 \apprle 6$, approaching the thermal form $\omega$
expected from the canonical ensemble.

\begin{figure}[bht]
\begin{center}
\includegraphics[scale=1.25]{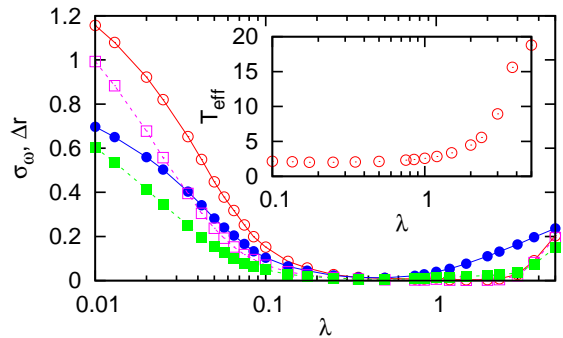}
\caption{Dependence on the initial state, $\Delta r$ (solid), and
distance from the thermal state, $\sigma_{\omega}$ (hollow), as a
function of coupling $\lambda$, at composite energies $E_0 = -2$
(circle) and $1.77$ (square).  Inset: effective temperature
$T_{\text{eff}}$ for $E_0 =-2$.}
\label{Teff}
\end{center}
\end{figure}

Next, we establish the range of the coupling $\lambda$ over which the
system forgets its initial state and thermalises. We expect the system
to retain memory of the initial state at weak coupling ($\lambda
\ll1$). Moreover, for $\lambda\gg 1$, the eigenspectrum becomes
significantly altered by the coupling, splitting into several bands
and we see oscillations.  This is a feature of the projection of the
initial state on the strongly-coupled link. Therefore, we expect that
the loss of memory of the initial state and thermalisation are
possible only in a range of intermediate couplings.  To quantify this,
we calculated the root-mean-square variation in diagonal RDM elements
due to using different initial subsystem states: $\Delta r =
\frac{1}{2} \sum_s [{\expectation{{r_{ss}}^2} -
\expectation{{r_{ss}}}^2}]^{\frac{1}{2}}$, with $\expectation{\ldots}$
averaging over all $16$ initial states in the subsystem Fock basis
(\emph{i.e.}, eigenstates at $J=0$).  A small $\Delta r$ indicates
memory loss.  We have also measured the closeness to the thermal state
$\omega$ using $\sigma_{\omega} = \frac{1}{2} \sum_s
\expectation{\left| {r_{ss}} - \omega_{ss} \right|}$. We see from
Fig.~\ref{Teff} that memory loss and thermalisation occur in the
intermediate range $\lambdathermal\apprle \lambda \apprle 3$ with
crossover value $\lambdathermal \simeq 0.1$ at $E_0=-2$ and $1.77$.

We also find that the relative probabilities of different states in
the $n_s=2$, $s^z_s=0$ sector fit a Boltzmann form: $\log r_{ss} =
-\varepsilon_s/T_{\text{eff}} +\text{const}$. For states near the
centre of the eigenspectrum ($E_0\simeq 1.77$), the effective
temperature $\Teff$ is infinite. At $E_0=-2$, we find
$T_{\text{eff}}\simeq 2$ up to $\lambda\sim 2$ (Fig.~\ref{Teff}
inset).  We estimate the chemical potential to be $2J\simeq 2$ so
that, unlike in previous work, we see thermalisation at temperatures
down to quantum degeneracy.

We note that these thermalised systems are surprisingly small.
Popescu \emph{et al.}~\cite{popescu1} give an estimate of the number
$d_R$ of composite-system eigenstates spanned by the initial state
sufficient for thermalisation --- if the probability that
$\sigma_{\omega}>Y=0.1$ is at least as small as $X=0.01$, then $d_R 
> (9\pi^3/2Y^2)\ln(2/X)
\simeq 70000$. This is almost two orders of magnitude larger than the
number of states ($\gtrsim \delta_B/\Delta$) spanned by our initial
state which is as low as 950. Moreover, we find
thermalisation at $U=J$ for smaller systems than at $U\ll J$.  We
believe strong inelastic scattering in the \emph{interacting} bath enables
efficient thermalisation at $U=J$ 
when system size is larger than the inelastic scattering length
($\propto J^2/U^2$ for small $U/J$).

\begin{figure}[htb]
\begin{center}
\includegraphics[scale=1.2]{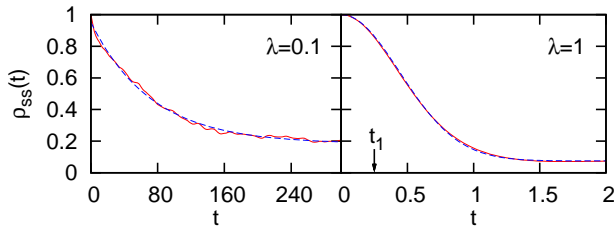}
\caption{The RDM element $\rho_{ss}$ as a function of time $t$ with $s =
\ket{\!\uparrow,\uparrow}$ for the initial state~\eqref{eq:windowstate}
at total energy $E_0 = -2$ ($\Teff\simeq 2$) with $s_i = s$. Left: coupling
$\lambda=0.1$ with exponential fit (dashed). Right: 
$\lambda=1$ with Gaussian fit (dashed).}
\label{shapes}
\end{center}
\end{figure}

\emph{Time Evolution.} Having established the coupling range for
thermalisation for model~\eqref{eq:2linkH}, we will now discuss our
main results for the temporal relaxation towards the steady
state. Fig.~\ref{shapes} shows examples of the time evolution of the
diagonal RDM element $\rho_{ss}(t)$ with $s=s_i$ for
two coupling strengths. 
The system is again prepared in the product
state~\eqref{eq:windowstate} with
$\ket{s_i}_S=\ket{\!\uparrow,\uparrow}_S$.  These results are computed
for energy $E_0= -2$. We do not
expect our results to depend strongly on $E_0$ unless 
the system is close to a strongly correlated ground state.

We find qualitatively different relaxation behaviour for perturbative
and non-perturbative couplings (Fig.~\ref{shapes}). 
Whereas the RDM relaxes towards the steady state exponentially in time at weak
coupling ($\lambda<\lambdaexp$), the relaxation follows a Gaussian
form at larger coupling ($\lambda>\lambdaGauss$). Interestingly, this
Gaussian regime covers the coupling range where the system
thermalises.

We can understand our results at short times or weak coupling.  At
short times, we can approximate $\ket{\Psi(t)}\simeq (1 -
iHt)\ket{\Psi(0)}$.  It can be shown (and our numerics agree) that the element
$\rho_{s_is_i}(t) \simeq 1-\Gammashort^2 t^2$ for $t< t_1 = 1/\max
(E_{s b} - E_0)$, with $\Gammashort = \lambda [\sum_{s\neq s_i,b}
|\bra{s b} V \ket{\Psi(0)}|^2]^{1/2} $.  The maximum energy difference
between states coupled by hopping ($V$) is of the order of
the single-particle bandwidth $4J$ and so $t_1 \simeq 1/4$.

\begin{figure}[thb]
\begin{center}
\includegraphics[scale=1.3]{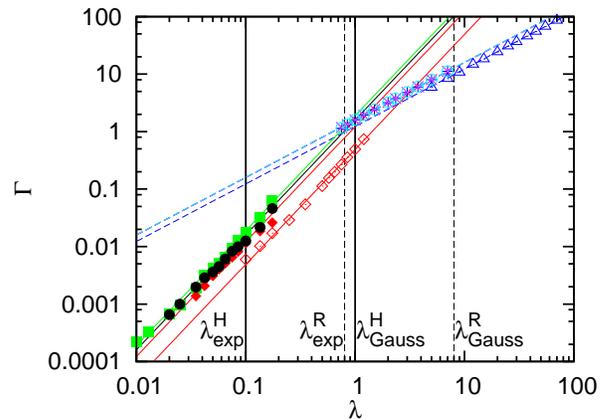}
\caption{(Colour online) Rates of decay of $\rdm_{s_is_i}$ for the
subsystem state $s_i = \ket{\!\uparrow,\uparrow}$ with initial
state~\eqref{eq:windowstate} at $E_0=-2$
($\CIRCLE$/$+$) and 1.77 ($\blacksquare$/$\times$) and random
model at $E_0=-2$ ($\Circle$/$\triangle$).  Weak coupling/exponential
decay: $-d\rho_{s_is_i}/dt|_{t=0}$ at short times found from
exponential fits ($\CIRCLE$,$\blacksquare$,$\Circle$) agree
with Fermi golden rule prediction, $\gammaFGR$ (solid lines,
gradient 2). Moderate coupling/Gaussian decay: fit to Gaussian with
rate $\Gamma$ ($\times$,$+$,$\triangle$) agrees with $\Gammashort$
(dashed lines, gradient 1). Vertical lines mark estimates of the
crossover values $\lambdaexp$ and $\lambdaGauss$
($\lambda^{\text{H/R}}$ for Hubbard/random models).  Data in the
crossover region are rates obtained from attempted fits to either
form.  }
\label{rate}
\end{center}
\end{figure}

At weak coupling, we can go beyond $t_1$ by treating the coupling $\lambda V$
as a perturbation to the uncoupled Hamiltonian $H_S +
H_B$. It is readily shown that, to leading order in $\lambda$, the
RDM element corresponding to a subsystem state $s\neq s_i$ is
approximated by $\rdmperturb(t)$:
\begin{equation}
\rdmperturb_{ss}(t) \!  =\! \frac{4\lambda^2}{B}\!\sum_{b}  
\!\left|  \sum_{b_i=b_l}^{b_u}\!
\frac{\sin[(E_{s b}\! -\!E_{s_i b_i})\frac{t}{2}]}{E_{s b} - E_{s_i b_i}} 
\bra{s b} V \ket{s_i b_i} 
\right|^2
\label{eq:FGR}
\end{equation}
after the composite system is prepared in the
state~\eqref{eq:windowstate}. The element $\rdmperturb_{s_is_i}$ is
most readily found by using $\text{Tr}
(\rdmperturb)=1$ to give $\rdmperturb_{s_i s_i}(t) = 1 - \sum_{s \ne
s_i} \rdmperturb_{ss}(t)$. This perturbation theory is valid until
time $t_2$ when $\rho_{s_is_i}$ has dropped significantly below unity.
For times between $t_1$ and $t_2$, eq.~\eqref{eq:FGR} follows the
Fermi golden rule (FGR): $\rdm_{s_i s_i}(t)$ decreases linearly in
time with $d\rdm_{s_is_i} /dt \propto -\lambda^2$.  Beyond the FGR
regime, we expect to see exponential decay (see, \emph{e.g.},
approximate Markovian schemes of the Lindblad type~\cite{BreuerPetruccione}) as is found in 
our data (Fig.~\ref{shapes} left) for $\lambda\apprle\lambdaexp =
0.1$. In our case, the initial state is not a bath eigenstate. This gives
small fluctuations on top of a simple linear-$t$ decay,
due to interference between terms in the inner sum in~\eqref{eq:FGR}.

We check in Fig.~\ref{rate} that the FGR prediction agrees
quantitatively with $d\rho_{s_is_i}/dt|_{t=0}$ for $E_0 = -2$ and
$1.77$, found from the parameters obtained for the exponential fit to
$\rho_{s_is_i}$ for $t>t_1$: $\rho_{s_is_i}(t) \sim A
e^{-(t-t_0)/\tau} + (1-A)$. The FGR rate, $\gammaFGR$, is found by
averaging~\eqref{eq:FGR} over a time $t$ between $t_1$ and $t_2$:
$-\gammaFGR t = \int_0^t [\rdmperturb_{s_i s_i}(t') - 1]dt'/t$.  This
procedure is needed for a non-zero level spacing $\Delta$. We point
out that the exponential fit fails at very weak coupling ($\lambda\sim
10^{-2}$) when the system barely relaxes.

We now consider larger couplings where $\lambda\sim O(1)$. Instead of
exponential relaxation, we find a good fit (Fig.~\ref{shapes} right)
to a Gaussian decay: $\rho_{s_is_i}(t) \sim C e^{-\Gamma^2 t^2} +
(1-C)$. This is seen for couplings $\lambda\apprge\lambdaGauss =1$.
The decay rate $\Gamma$ now increases linearly with $\lambda$ and
is as large as the bandwidth scale $1/t_1 \simeq 4$.  It appears
insensitive to energy $E_0$.  Interestingly, we see in Fig 4 that, in
the regime where $\Gamma t_1 \sim 1$, the decay rate $\Gamma$ is well
approximated by $\Gamma_1$ from the short-time expansion, which
suggests $\Gamma = \Gamma_1/C^{\frac{1}{2}}$. (In our data,
$0.97<C^{\frac{1}{2}}<1$.)  In other words, perturbation theory gives
the early-time precursor to the full Gaussian form. This suggests the
interpretation that the time interval of validity of the Fermi golden
rule ($t_1<t<t_2$) narrows and vanishes as $\lambda$ increases to
$\lambdaGauss$.  For coupling range $\lambdaexp<\lambda<\lambdaGauss$,
the behaviour is less clear cut --- the decay starts as a Gaussian but
becomes exponential at later times. The amplitude of this exponential
tail decreases with increasing $\lambda$, becoming negligible as
$\lambda$ reaches $\lambdaGauss$.

\emph{Random Couplings.}~To verify
that the two regimes of relaxation are not
specific to our model Hamiltonian, we proceeded to study an
alternative model where the subsystem-bath coupling $V$ is replaced
by a random Hermitian matrix $W$ which still respects the
conservation of the global particle number $N$ and spin $S^z$.  Each
non-zero matrix element of $W$ is Gaussian distributed, with the
variance chosen such that $\text{Tr}
(W^2)=\text{Tr} (V^2)$. Thus, we can compare $H=H_S+H_B+\lambda V$
with $H=H_S+H_B+\lambda W$ with similar decay rates. In this model,
we expect $1/t_1$ to be of the order of the full bandwidth 
$\sim 20$ for $N=8$, $S^z=0$.
We find exponential relaxation at weak coupling,
$\lambda \apprle\lambdaexp = 0.8$, and we recover 
Gaussian relaxation  with a linear-$\lambda$
decay rate for $\lambda \apprge\lambdaGauss = 8$
(Fig.~\ref{rate}, hollow symbols). 
The crossover values,
$\lambdaexp$ and $\lambdaGauss$, occur at nominally higher couplings 
than for the Hubbard ring~\eqref{eq:2linkH}. They
become closer to the Hubbard-ring values if we 
mimic the structure of $V$ by restricting the 
states coupled by $W$: $\bra{s'b'}W\ket{sb}\neq 0$ only if
$|E_{s'b'}-E_{sb} |<4J$, the single-particle bandwidth.

\begin{figure}[thb]
\begin{center}
\includegraphics[scale=0.35]{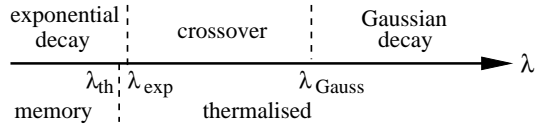}
 \caption{Schematic behaviour of the reduced density matrix $\rho(t)$
as a function of subsystem-bath coupling $\lambda$. Top: relaxation to steady state. Bottom: steady state of $\rho(t)$. 
Memory of initial state also occurs at large $\lambda$ in the Hubbard case.}
\label{schematic}
\end{center}
\end{figure}

We summarise our results in
Fig.~\ref{schematic}.  We have shown that a two-site subsystem of the
Hubbard model relaxes to steady states resembling canonical thermal
states, even for systems with a handful of sites and at quantum degenerate energies. 
This occurs at a non-perturbative coupling between the subsystem and bath, 
corresponding to nearly homogeneous systems.
In this regime, the reduced density matrix
$\rdm(t)$ displays Gaussian relaxation to the thermal state, with a
decay rate $\Gamma$ linear in the coupling
$\lambda$. This contrasts sharply with the perturbative regime where
$\rdm(t)$ exhibits an exponential relaxation with a
$\lambda^2$ decay rate. We believe that the
Gaussian relaxation to thermalisation is a
generic feature of closed nanoscale systems, 
as is supported by our results for random Hamiltonians. 


Finally, we note that it can be shown that $\Gamma_1 t_1 \sim \lambda J t_1 \sim \lambda$ irrespective of system size. 
The subsystem thermalises on the time scale of a few hops between the subsystem and the bath, 
by inelastic collisions of the fermions within this timescale. This should be insensitive to system size
for systems larger than the inelastic scattering length. 
Therefore, we speculate that the observed Gaussian relaxation should remain for large systems.

We are grateful to Miguel Cazalilla for useful discussions.
We wish to thank Imperial College HPC for computing resources as well
as EPSRC for financial support. 
\bibliographystyle{apsrev4-1}

\end{document}